\documentclass[a4paper, 11pt]{article}
\pdfoutput=1
\usepackage{array}
\usepackage{tikz}
\usepackage{booktabs}
\usepackage{gensymb}
\usepackage{jinstpub}
\usepackage{float}
\usepackage{lineno}
\usepackage{comment}
\usepackage[colorinlistoftodos]{todonotes}

\graphicspath{{./Figures/}}
\proceeding{LIDINE 2017: Light Detection in Noble Elements\\
22-24 September 2017\\
SLAC National Accelerator Laboratory}  

\title{On the Evaluation of Silicon Photomultipliers for use as Photosensors in Liquid Xenon Detectors}

\author{Benjamin Godfrey,} \author{Tyler Anderson,} \author{Earl Breedon,} \author{Jacob Cutter,} \author{Navneet Dhaliwal,} \author{Olivia Dalager,}  \author{Seth Hillbrand,}  \author{Michael Irving,} \author{Aaron Manalaysay,} \author{Juan Montoya,} \author{James Morad,} \author{Christian Neher,} \author{Dustin Stolp,} \author{Mani Tripathi,} \author{Ryan Wilson}
\affiliation{University of California, Davis,\\
  One Shields Avenue, Davis, CA}
  
\emailAdd{bpgodfrey@ucdavis.edu}
  
\abstract{Silicon photomultipliers (SiPMs) are potential solid-state alternatives to traditional photomultiplier tubes (PMTs) for single-photon detection. In this paper, we report on evaluating SensL MicroFC-10035-SMT SiPMs for their suitability as PMT alternatives. We successfully operated these devices in a liquid-xenon detector, which demonstrates that SiPMs can be used in noble element time projection chambers as photosensors. The devices were also cooled down to 170\,K to observe dark count dependence on temperature. No dependencies on the direction of an applied 3.2\,kV/cm electric field were observed with respect to dark-count rate, gain, or photon detection efficiency.}


\begin{document}
\maketitle
\flushbottom

\section{Introduction}

Dual-phase noble element time projection chambers (TPCs) are a commonly used technology  (\cite{DarksideOverview}, \cite{LUXTotalResults}, \cite{XENON100Overview}, \cite{ArDMOverview}, \cite{LZOverview}) for the direct detection of weakly interacting massive particles (WIMPs). They record primary, prompt scintillation from interaction with the liquid target (S1) and secondary, delayed luminescence from ionized electrons drifted up in the liquid and extracted into the gas phase (S2)\cite{TPCOverview}. Photosensors positioned around the noble-element volume are then used to detect the S1 and S2 signals. Traditionally, these sensors have been photomultiplier tubes (PMTs), which are typically deployed on the top and bottom surfaces of a cylindrical liquid volume. The walls of the volume are constructed using a high-reflectivity material, such as polytetrafluorethylene (PTFE).
Recent major detectors have achieved light collection efficiencies (LCE) in the 30-40\% range (\cite{LCE_LUX}, \cite{LCE_Xenon1T}), resulting in a photon detection efficiency of 9-12\% due to the ~30\% quantum efficiency.

One approach towards increasing the LCE of noble-element TPCs would be to replace PTFE reflectors with an active detection element. However, PMTs are not suitable because they will not operate in the high electric field in that region of the TPC. Solid-state substitutes to PMTs are silicon photomultipliers (SiPMs), which are  arrays of avalanche photodiodes operating in Geiger mode. Their small size, low operating voltage, single-photoelectron sensitivity, and continually decreasing cost make them competitive alternatives for PMTs~\cite{SiPMCSeries}. 

\begin{figure}[h]
\centering
\includegraphics[width=3.1in]{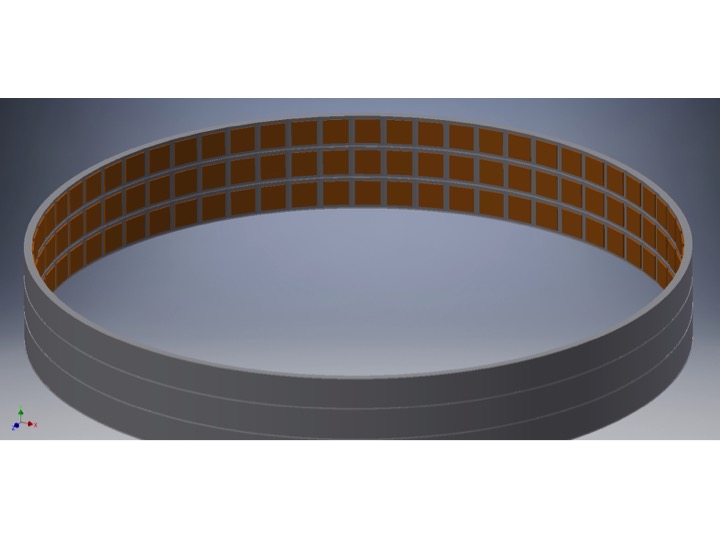}
\includegraphics[width=2.4in]{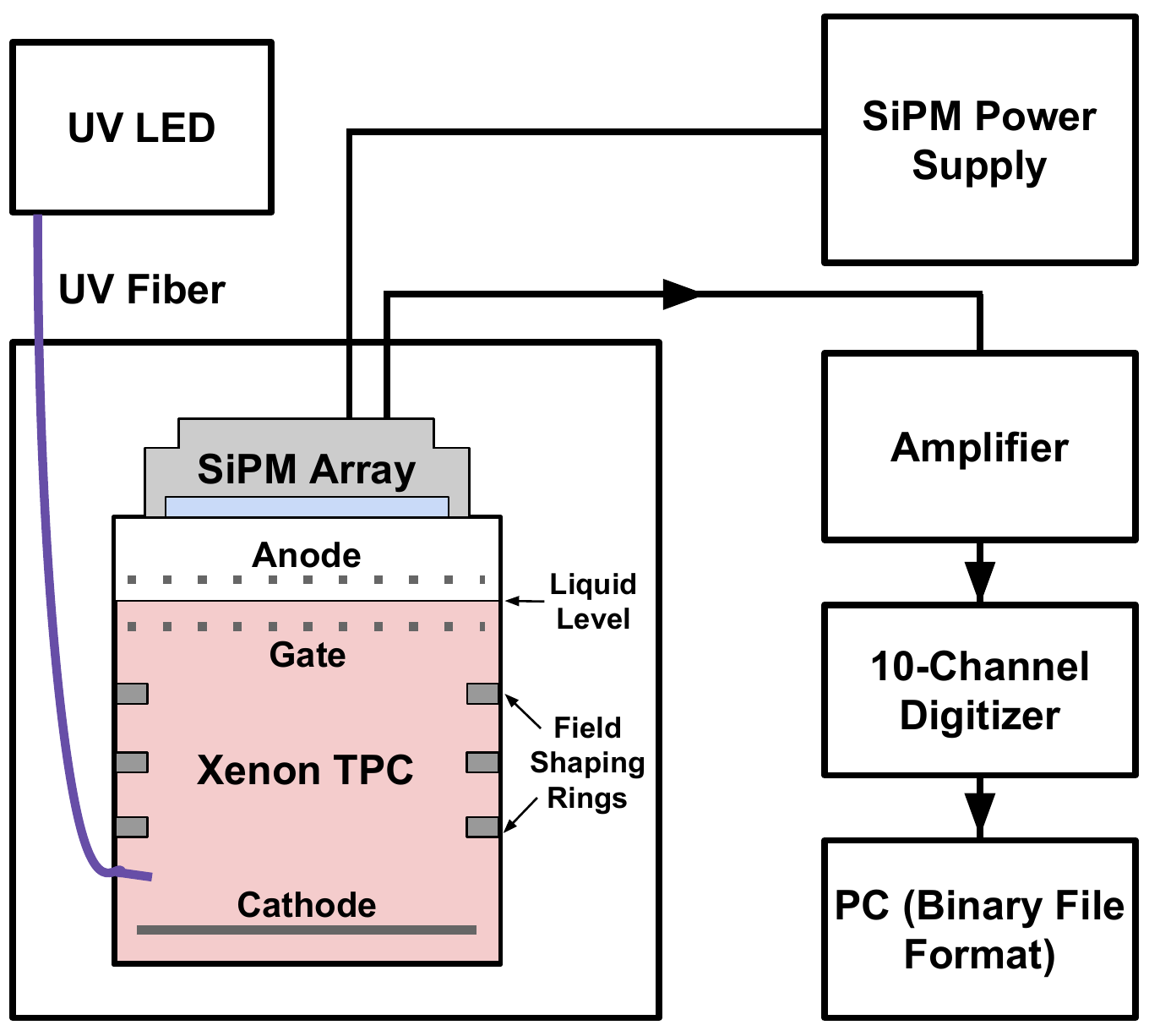}
\caption{(Left) A schematic depiction of a section of the TPC volume. SiPMs, tiled on field-shaping rings that form the cylindrical shell of a TPC chamber, allow for increased light detection coverage. (Right) A schematic diagram of the DAX TPC instrumented with the SiPM array.}
\label{fig:ExampleTPC}
\end{figure}

In our proposed design, SiPMs are mounted on the field shaping rings forming a cylindrical shell, as shown in  Figure ~\ref{fig:ExampleTPC} (Left). This design, however, assumes both operation of these devices in high electric fields and at the low temperatures of the liquid noble-element. Furthermore, the SiPMs must have a non-zero quantum efficiency (QE) for photo-electric conversion of photons in the UV wavelengths of interest. Previously, we reported (\cite{DustyPaper}) on using a wavelength shifting material in conjunction with SiPMs to achieve this. 

Herein, we report on investigations of the suitability of SiPMs to be used in the proposed design concept by operating an array of SiPMs in a liquid xenon detector.  In a separate experiment, we measured dark count rates as a function of temperature down to 170 K. These results extend work that has been done characterizing SiPMs in low temperature cryogenic environments (\cite{OldAprile} \cite{Catalanotti}, \cite{Lightfoot}) to SensL's C-series SiPMs. In a third setup, we measured  dark counts, gain, and photon detection efficiency (PDE) of Sensl MicroFC-10035-SMT SiPMs in a constant electric field of 3.2 kV/cm, as a function of the field direction. 

\section{Operating a SiPM Array in Liquid Xenon}

\begin{figure}
\centering
\includegraphics[width=1.25in]{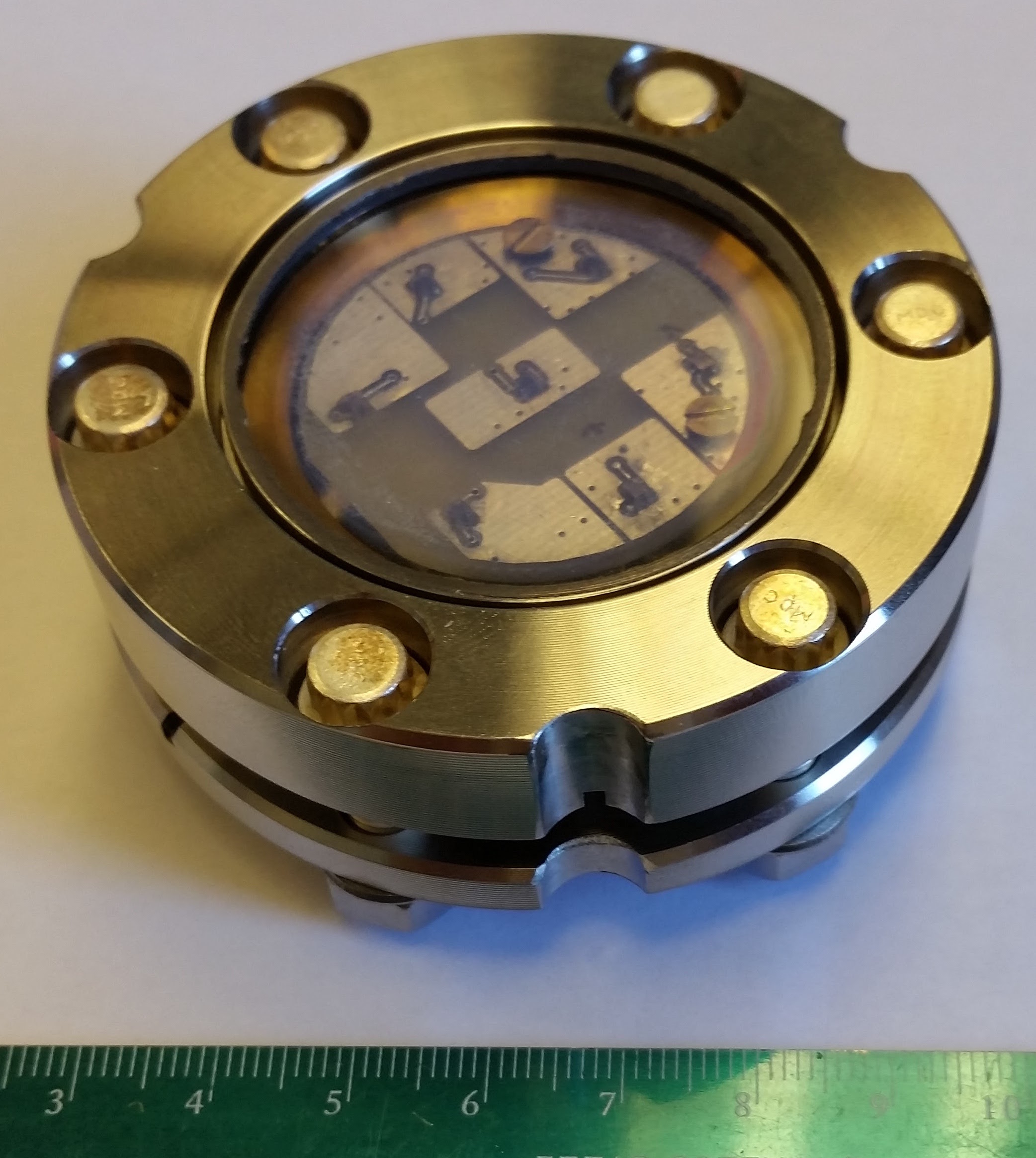}
\includegraphics[width=1.5in]{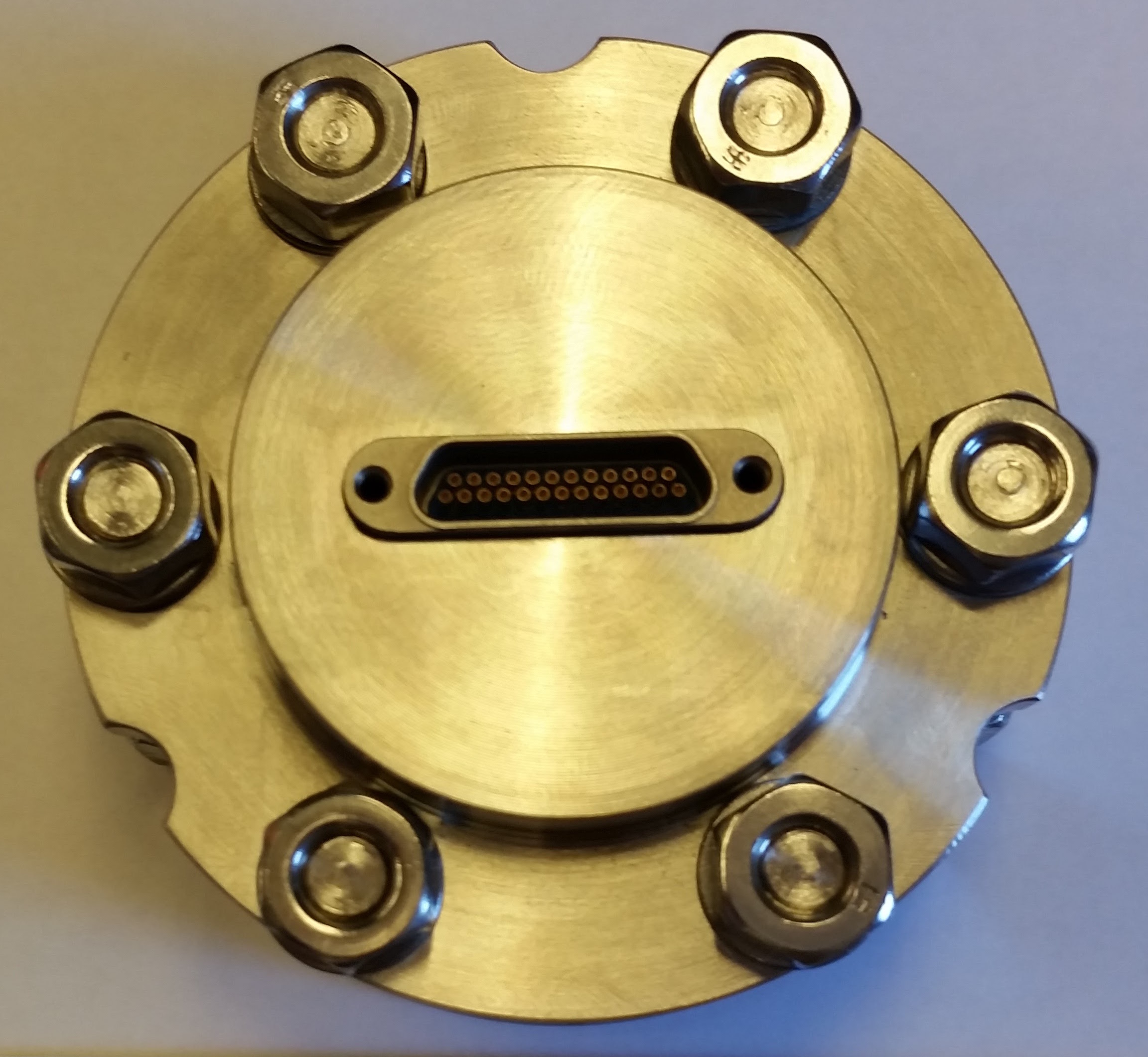}
\caption{Photographs of the front (Left)  and back (Right) of the SiPM flange assembly. The front face shows the fused silica window, with TPB deposited on the inside, and the array of seven SiPMs.}
\label{fig:SIPM_Flange}
\end{figure}

A schematic diagram of the experimental setup, involving the Davis Xenon (DAX) detector, is shown in Figure~\ref{fig:ExampleTPC} (Right). DAX is designed to be a flexible detector that can be configured in various modes to serve as a liquid-phase or a dual-phase xenon detector.  For these measurements, the cathode consisted of a copper plate, and the anode was made up of a patterned foil. An array of seven SiPMs, as shown in Figure \ref{fig:SIPM_Flange}, was deployed at the top. The SiPMs were mounted on a PCB, which was sandwiched inside two flanges - the front facing flange had a fused silica window, which was coated with a UV-transparent 600 nm thick layer of tetraphenyl-butadiene (TPB) supplied by Sigma-Aldrich, and the back face had a DB-25 connector used for providing bias voltages and bringing out the standard output (SOUT) signals from the SiPMs. The thickness of TPB was optimized in our earlier work \cite{DustyPaper}.  

A custom power supply, with an adjustable range of 27-32V per channel, provided individually optimized bias voltages for each SiPM. The SiPM signals were amplified using custom shaping amplifiers and digitized using the DDC-10 data acquisition system \cite{DDC-10}. An optical fiber was introduced into the volume to provide UV light pulses from an LED. This was used for initial testing of the SiPM array to prove responsiveness of the array to UV light but was not used for calibrations. The DDC-10 system was operated in a self-triggered mode, resulting in the capture of an 80 $\mu$s window, digitized to 14 bits at 100 Msps, corresponding to 8,000 ADC samples centered at the trigger time stamp. Pulse areas were determined using a custom pulse finding algorithm \cite{DustyPaper}.

\begin{figure}
\centering
\includegraphics[width=2.9in]{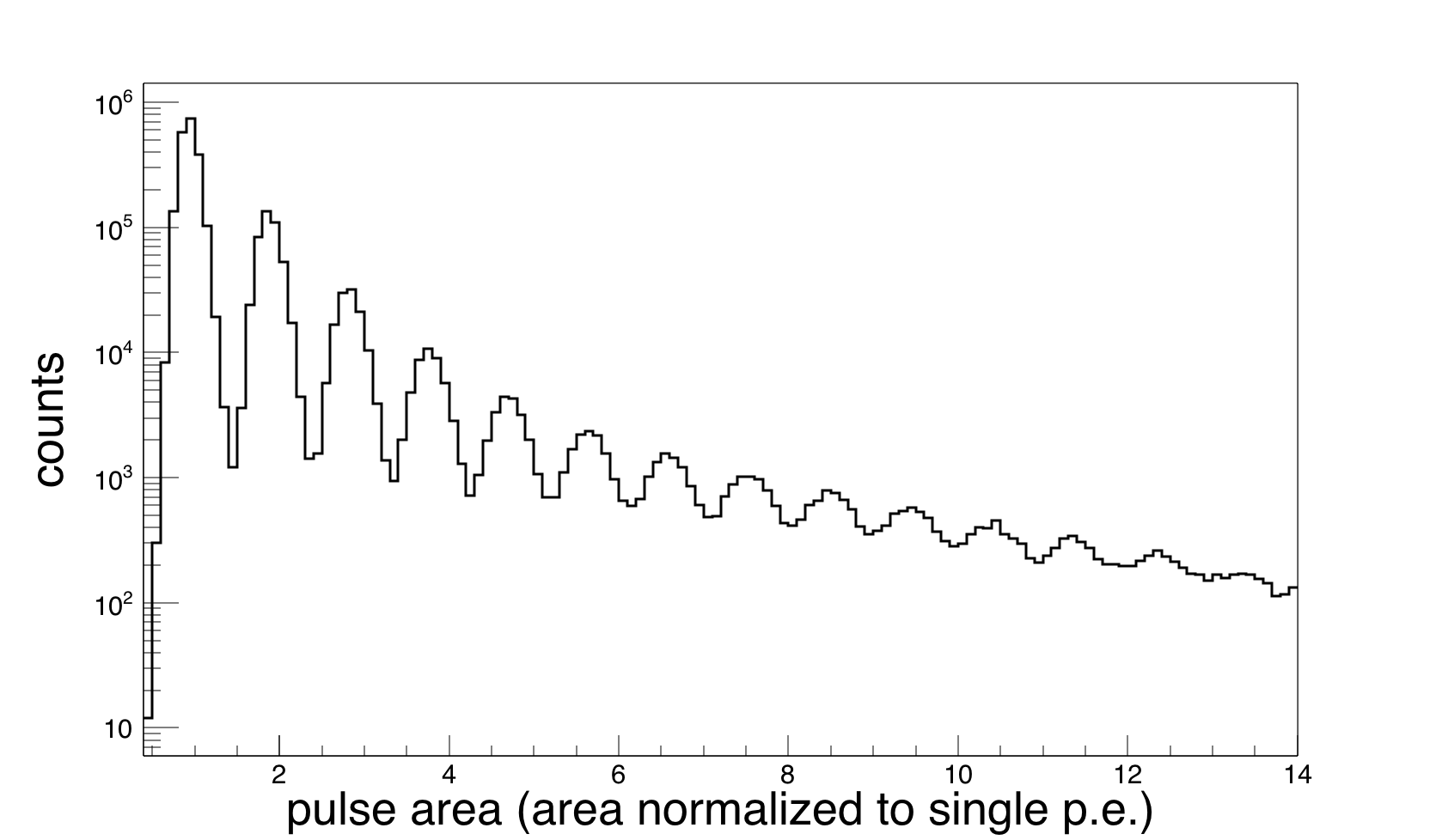}
\includegraphics[width=2.9in]{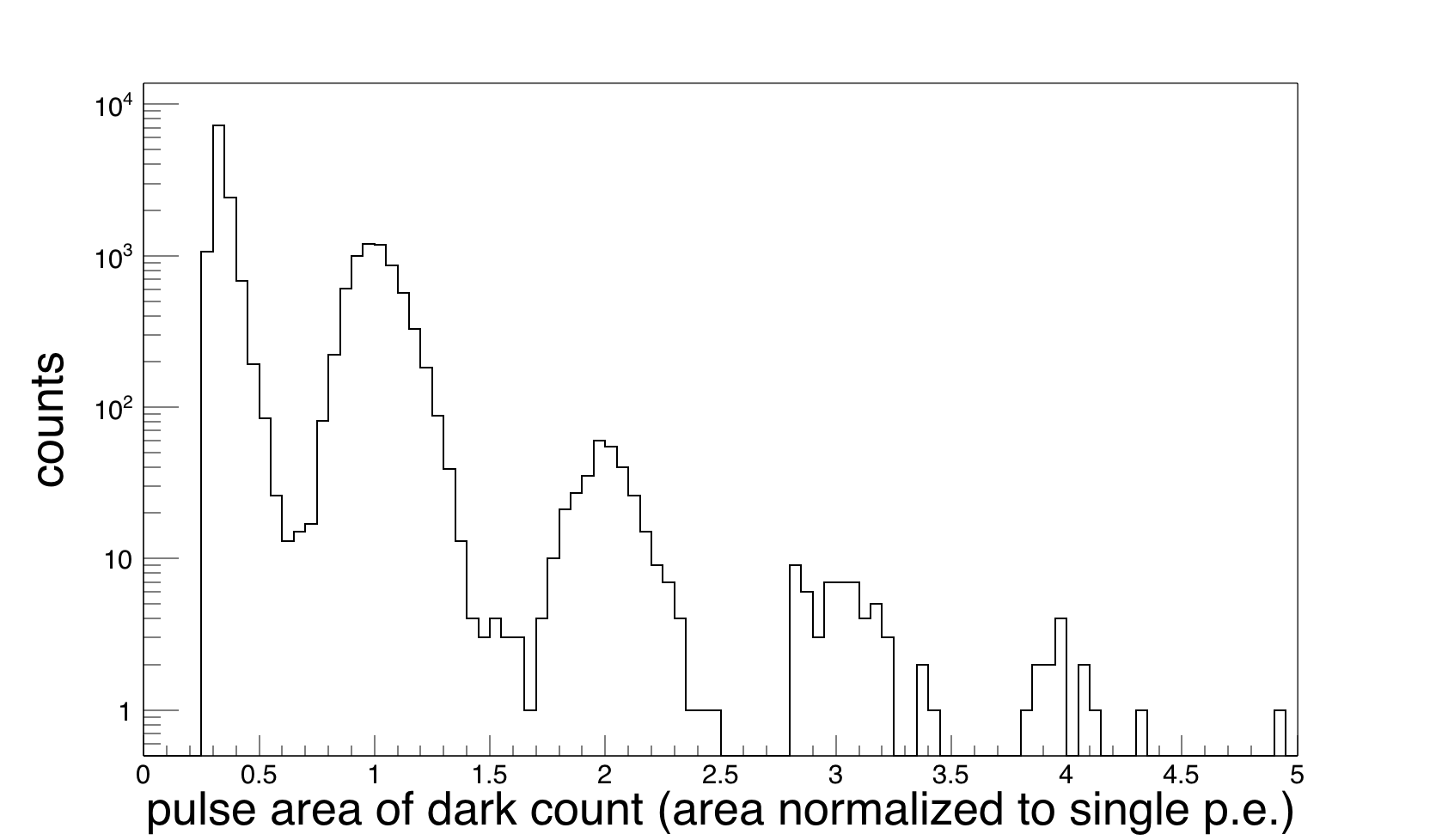}
\caption{Histograms of pulse areas collected from all SiPM channels. (Left) Response to $\alpha$-particles created from the decay of Po-210, and (Right) Dark counts measured in all channels.}
\label{fig:SIPM_pulses}
\end{figure}

\begin{figure}
\centering
\includegraphics[width=3.3in]{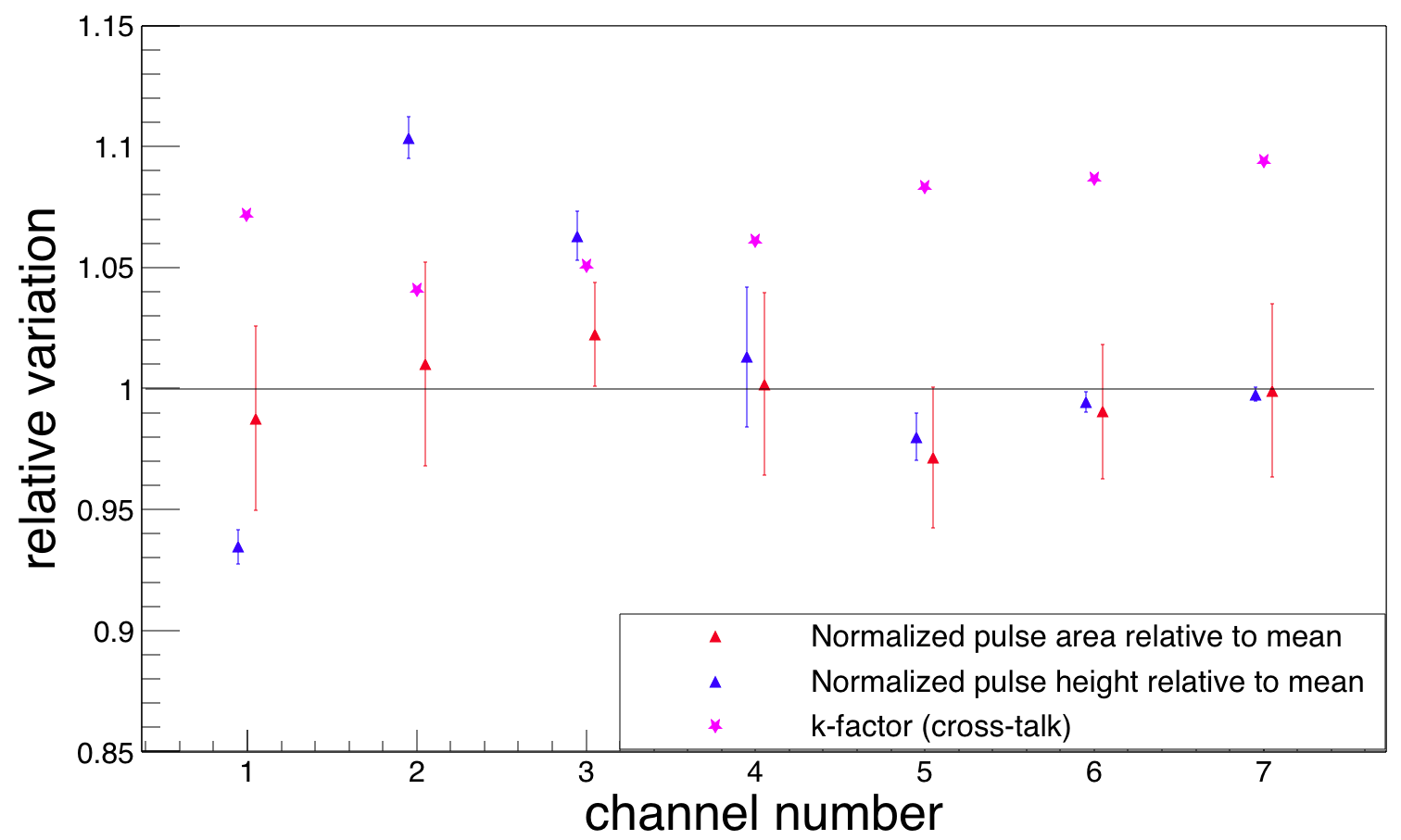}
\includegraphics[width=2.3in]{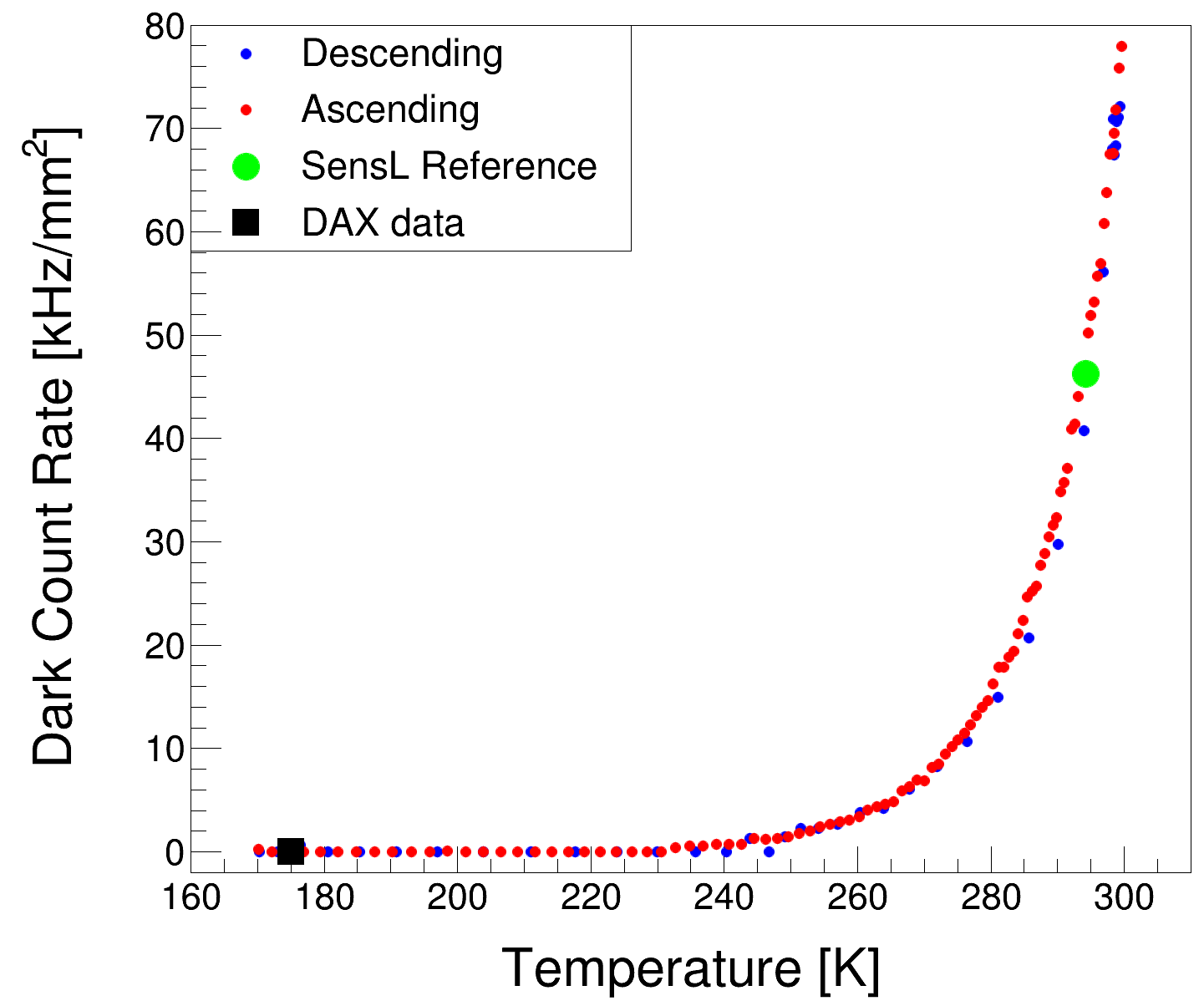}
\caption{(Left) Relative variations in gain parameters and cross-talk, as described in the text, for the seven SiPM channels. (Right) Dark count dependence on temperature for both cooling (descending) and heating (ascending) curves.  The point labeled as \textit{SensL Reference} is obtained from \cite{CSeriesReference}. Below 240 K the dark count rate is averaged to be 19.1 Hz $\pm$ 1.2 Hz. The point labeled as \textit{DAX data} is the measurement made in the previous section.}
\label{fig:SIPM_calibrations}
\end{figure}

Figure \ref{fig:SIPM_pulses} (left) shows a histogram of S1 pulse areas recorded by all of the SiPMs in response to $\alpha$-particles, created from the decay of a Po-210 source, deposited on the cathode \cite{JacobPaper}. The multiple photoelectron peaks are clearly visible. However, the 7 mm$^2$ area of the SIPMs was insufficient to collect enough photons to reconstruct the energy of the incident $\alpha$-particle.  This study determines the ability of a SiPM-TPB combination to detect 175 nm scintillation photons in a liquid xenon detector.  
These data were also used to characterize dark counts and cross-talk rate variations for the seven SiPM channels.  We searched for pulses in the digitized sample \textit{preceding} the trigger to accumulate a sample of dark counts per channel. Figure \ref{fig:SIPM_calibrations} (Left) shows the relative variations in response \textemdash pulse areas and pulse heights, normalized to the mean of the seven channels. Also shown is the variation in k-factor, defined as the fraction of pulses with an area greater than one photoelectron out of the total number of pulses. The dark count rate, averaged over the seven channels, was 25$\pm$5 Hz/$\text{mm}^{2}$. This corresponds to a dark current of 0.59$\pm$0.12 nA \cite{SensLDatasheet}.

In order to measure dark counts versus temperature, an unmodified SensL MicroFC-SMA-10035 evaluation board was placed inside a copper vessel that was filled with thermally conductive foil. 
This vessel, in turn, was placed inside a Styrofoam container, in which liquid nitrogen was poured, to provide cooling. The SiPM was reverse biased at 28.4 V, and the SOUT signal was fed into a wide bandwidth amplifier, which was then digitized by a DDC-10 digitizer. Temperature was monitored and controlled using a cryogenic temperature controller connected to a silicone-rubber heat sheet deployed inside the vessel.

Data acquisition was split into two parts: Cooling and heating. This was done in order to check for consistency and remove any systematic error in the setup. 
Temperature was monitored continuously using the temperature controller. Data were recorded in 80 ms blocks, corresponding to $8*10^{6}$ ADC samples, by the DDC-10, which was externally triggered by a 5 kHz square wave. 
Dark count peaks were counted in software using the same pulse finding script described previously. All temperatures recorded during each 80 ms window were averaged to give a mean temperature associated with the number of dark count peaks. 

Figure \ref{fig:SIPM_calibrations} (Right) shows the measured dark count rate as a function of temperature. Heating and cooling curves reliably lie on top of one another confirming consistency of the setup. The dark count rate falls sharply from 300 K to about 260 K, and below 240 K it is nearly constant with an average of 19.1 Hz $\pm$ 1.2 Hz. 

\section{High Electric Field Tests} 
\label{section:ElectricField}

\begin{figure}[h!]
\centering
\includegraphics[width=3.0in]{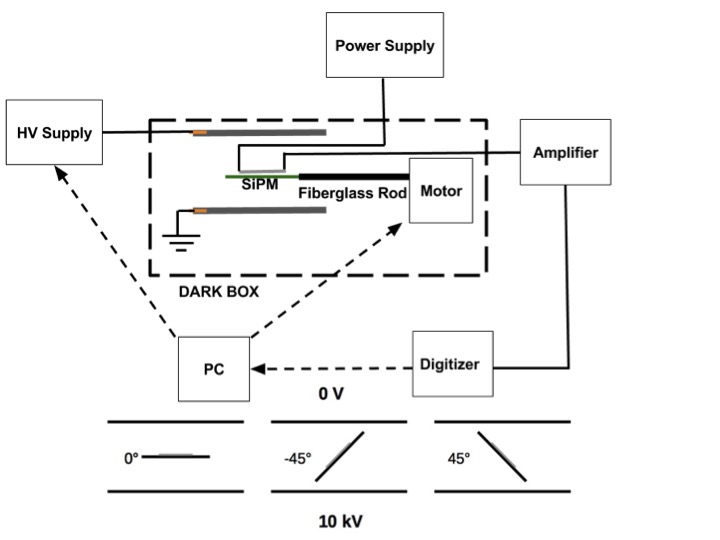}
\includegraphics[width=2.8in]{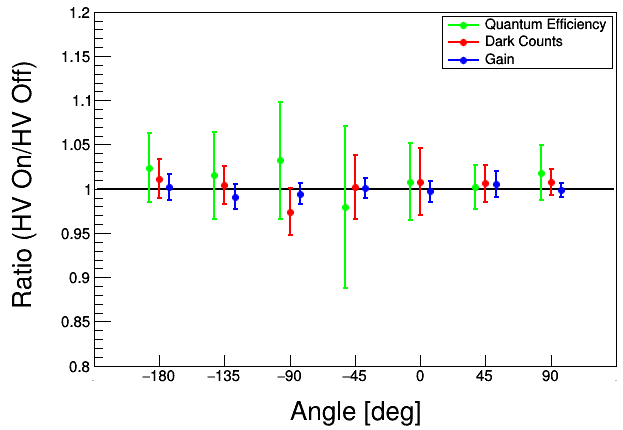}
\caption{(Left) A schematic diagram of the setup for testing dark counts, gain, and PDE of SensL MicroFC-10035-SMT SiPMs in high electric fields with definition of the servo angles underneath. At $0^\circ$ the SiPM faces the top plate and at $-180^\circ$ the SiPM faces the bottom plate. $-180^\circ$ and $+180^\circ$ refer to the same angle, but the servo is unable to go beyond $+115^\circ$, so $+180^\circ$ is not discussed. (Right) Relative QE, dark counts, and gain as a function of angle.}
\label{fig:ElectricFieldSetup}
\end{figure}

An outline of the setup is detailed in Figure~\ref{fig:ElectricFieldSetup}. A high electric field was created between two circular, aluminum plates, approximately 25 cm in diameter and 1/16"  thick, held approximately 3.1 cm apart. High voltage was supplied from a PS365 programmable power supply from Stanford Research Systems.

A SiPM mounted on a custom PCB was placed at the end of a 15 cm long G-10 fiberglass rod. The board was held near the center of the two plates in the uniform field region. The rod was connected to a Dynamixel AX-12A actuator, computer controlled through a serial link. Throughout these tests, the SiPM was reverse biased at a fixed 28.4 V, and was electrically isolated from the HV power supply such that the entire board was allowed to float to any voltage.

To evaluate the performance of the SiPM in arbitrarily oriented electric fields, the board, attached to the actuator's fiberglass rotor, was positioned at the fixed angles of $-180\degree$, $-135\degree$, $-90\degree$, $-45\degree$, $0\degree$, $45\degree$, and $90\degree$. Positive and negative angles are defined in Figure~\ref{fig:ElectricFieldSetup}.  The upper and lower bounds were set by hardware limitations of the actuator. Data collection steps were repeated with the electric field on and off.  As before, the SOUT signal of the SiPM was fed into a shaping amplifier and digitized by a DDC-10 in 80 ms windows. Dark counts were collected eight times for each angle setting, which were then averaged to determine the rate. The standard deviation among the eight sets was considered to be a systematic error likely due variations in the illumination from the LED light source.

A relative  measurement for various quantities was computed as the ratio of field-on and field-off data.  As seen in Figure \ref{fig:SIPM_pulses} (Right), the first peak (centered around 1 p.e.) is due to single photoelectrons, while the second is mostly due to optical crosstalk, which is common in SiPMs ~\cite{CSeriesReference}. A Gaussian fit was performed on both of these peaks, and the gain is defined as the difference in the mean values as given in \cite{SensLOverview}.

The QE of the SiPM is defined as the probability that an incident photon will produce a  pulse from one of its microcells \cite{SensLOverview}.  Since the number of incident photons on the SiPM when the HV is on is the same as when it is off, the relative QE is given by Equation~\ref{eq:RelativeQE}. For the QE measurement, two Arduino-controlled LEDs were added to the dark box to ensure ample illumination of the SiPM at all angles.

\begin{equation}
\textnormal{Relative QE} = \frac{\textnormal{Counts  with HV on and LED on - Counts with HV on and LED off}}{\textnormal{Counts with HV off and LED on - Counts with HV off and LED off}}
\label{eq:RelativeQE}
\end{equation}

Possible systematic error due to the effects of crosstalk and after-pulsing was determined to be negligible. Since only the number of peaks are being considered, crosstalk does not affect this ratio. Furthermore, since the number of counts due to after-pulsing is proportional to the total number of counts, any  possible effects cancel out when doing a relative measurement. 

For each angle, eight pulse-area histograms were created from which eight gains were recorded and averaged.   For the relative QE measurement, the SiPM was rotated through all angles with HV on and LED on, then HV on and LED off, then HV off and LED on, and, finally, HV off LED off. In addition, a five-minute delay after the LED was turned on, and a 30-second delay after the LED was turned off, were included to give the temperature of the dark box time to stabilize.

Figure~\ref{fig:ElectricFieldSetup} (Right) shows these relative measurements as a function of angle. There is no significant deviation from unity for all three ratios, thus demonstrating that there is no discernible effect of the electric field on these quantities.

\section{Conclusions}
We have observed that dark count rate, gain, and QE are independent of electric field strength up to 3.2 kV/cm. Combined with greatly reduced dark count rates at near-cryogenic temperatures, SiPMs show promise not only as alternative photon detectors for large-area coverage in noble element TPCs but also in other low-temperature, high electric field environments where sensitive photo-detection is required.


\section{Acknowledgments}

This material is based upon work supported by the Department of Energy National Nuclear Security Administration under Award Number: DE-NA0000979 through the Nuclear Science and Security Consortium.

This report was prepared as an account of work sponsored by an agency of the United States Government. Neither the United States Government nor any agency thereof, nor any of their employees, makes any warranty, express or limited, or assumes any legal liability or responsibility for the accuracy, completeness, or usefulness of any information, apparatus, product, or process disclosed, or represents that its use would not infringe privately owned rights. Reference herein to any specific commercial product, process, or service by trade name, trademark, manufacturer, or otherwise does not necessarily constitute or imply its endorsement, recommendation, or favoring by the United States Government or any agency thereof. The views and opinions of authors expressed herein do not necessarily state or reflect those of the United States Government or any agency thereof.

\newpage
\clearpage
\bibliographystyle{unsrt}
\bibliography{References}
\end{document}